\documentclass{ws-ijmpd}

\begin{document}

\markboth{G. Chincarini, R. Margutti}
{Swift highlights and flares}

%
\catchline{}{}{}{}{}
%

\title{Swift highlights and flares\\(Back to the drawing board?)}

\author{G. Chincarini,$^{1,2}$ R. Margutti,$^{1,2}$}

\address{
$^{1}$ Dipartimento di Fisica ``G. Occhialini'', Universit\`a Statale di Milano Bicocca, Piazza della Scienza 3, I-20126 Milano, Italy.\\
$^{2}$ INAF - Osservatorio Astronomico di Brera, via Emilio Bianchi 46, I-23807 Merate (Lc), Italy.\\ 
E-mails: guido.chincarini@brera.inaf.it, raffaella.margutti@brera.inaf.it.
}

\maketitle

\begin{abstract}
Swift opened up a new era in the study of gamma-ray burst sources (GRB). Among a variety of discoveries made possible by Swift, 
here we focus on GRB\,090423, the event at z=8.2 which currently holds the record of the most distant celestial object ever caught 
by human instrumentation. This GRB allowed us to have a direct look at the early Universe. The central engine activity giving origin 
to the GRB emission  is also discussed starting from the observational findings of an updated GRB X-ray flares catalog. 
\end{abstract}


\section{Introduction}
Each time we build and use innovative state of the art instrumentation we expect to detect new phenomena and details that allows a better understanding of the physics of the objects we are observing. In the case of the Swift mission\cite{Gehrels04} we were able to build on the experience obtained by Beppo-Sax and on the theoretical developments that helped the design allowing targeting critical points of the GRB physics. The result was a multi-wavelength mission with very fast pointing, accurate astrometry of the transients and real time analysis and communication of the detected targets.\\

Out of the variegated science and related high lights we have recently published, here we will discuss GRB\,090423, a burst holding 
the high z record for any celestial object so far discovered, and   illustrate some of the results we obtained with the new sample of 
GRB flares. In other words we will look into the early Universe and perhaps into the activity intimately related to the central 
engine.\\

Nowadays we have evidence that the early star formation, pop III, is followed by the re-ionization (of HI) phase that likely terminated 
at $z \sim 6$. As it is well known even a small amount of neutral hydrogen would heavily absorbs at wavelengths smaller than Ly$_{\alpha}$(Gunn - Peterson effect) so that this region in which some neutral hydrogen still exist in the IGM is easily recognizable 
and indeed it is a powerful tool to track down the various phases of the process as a function of redshift.\\
In the X-ray light curve of the afterglow (generally covering the temporal range of few hundreds second since trigger to a million second after that) we define as flare an event that manifests itself as a sudden change in the observed flux and lasting a time largely smaller than the light curve of the GRB itself. The cause of these later injections of energy is not yet known and while it may be caused
either by late shocks among shells emitted during the prompt emission or, more likely, by renewed activity of the central engine, they 
are certainly not due to external shock (it can be shown that the pulse width divided by the time of occurrence of the pulse is too small to be explained as due to external shock). But no matter what, a clear understanding of the observed emission will lead to the understanding of the mechanism at work and eventually lead to the understanding of the central engine. Any correlation among parameters 
of flares occurring at different time in different bursts must relate to a behavior of the central engine and may lead to its partial
understanding.

\section{The re-ionization epoch and GRB\,090423}
Following the formation of neutral and molecular hydrogen (the cosmic abundance is that determined by the primordial nucleosynthesis) 
the formation of the Pop III stars reionizes the intergalactic medium. Information about this important phase of the cosmic 
evolution comes from the objects observed at very high z.  High z objects (the search of high z objects has been an observational 
game since ever) can be detected using various techniques and various dedicated surveys, among these the fundamental Hubble 
Space Telescope Ultra Deep Field survey and the SDSS have lead to fundamental results.  For $z > 5$, the drop out 
technique\cite{Steidel96} using multi filters optical observations is the least biased and likely gives the strongest indication 
for a spectroscopic follow up. One of the major goal justifying the search of high z galaxies is, in addition to the understanding of 
the formation and evolution of Pop III stars, the understanding of the sources that reionize the observations at that epoch (for a 
brief and clear description of the problem see Ref.\,\refcite{Salvaterra10} and references therein). The most distant galaxy has been detected at z = 6.96\cite{Iye06} while the most distant AGN has been detected at $z \sim 6.43$\cite{Willott07}. Photometric 
indications (these galaxies and AGN are too faint to get a spectrum even with the very large telescopes) exist of objects with 
$7 < z < 10$; what is really needed is the spectrum in order to have not only a certain identification but also the possibility 
to measure continuum and lines to estimate the population and the metal abundance.\\

In the last few years, and this was one of the goals of the Swift mission, the GRBs came into the game. Their very large luminosity 
makes them visible and detectable at very high z and the featureless power law spectrum of the afterglow make of them an ideal beacon 
to identify the absorption lines of the ISM of host galaxy and the characteristics of the IGM with the advantage on the AGN of not 
being affected by the proximity effect. There is always a price to pay and that is due to the rapid decay of the luminosity so that 
we must target them soon after the detection of the prompt emission. To this end the technology is at hand however; we need to build 
3 to 4 meters aperture robotic telescopes with state of the art focal plane 
instruments\cite{Vitali10,Chincarini10}.\\

Swift detected three objects for which the optical follow up evidenced through their spectra very high z objects: 
GRB\,050904 at z = 6.29\cite{Kawai06}, GRB\,080913 at z=6.7\cite{Greiner08} and 
GRB\,090423 at z = 8.2\cite{Salvaterra09,Tanvir09}.  The latter hold the record for any celestial object so far observed.\\

\begin{figure}
\begin{center}
\includegraphics[width=\hsize]{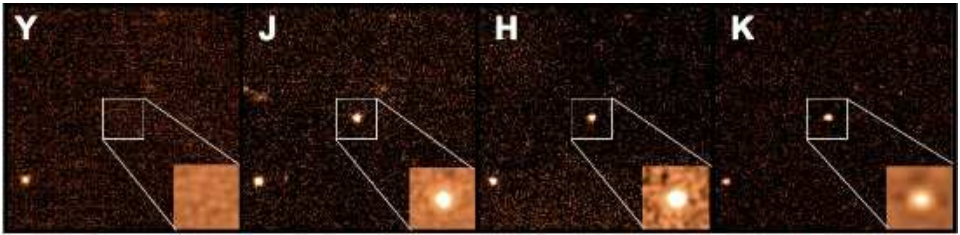}
\caption{These images obtained with the UKIRT telescope in Hawaii clearly indicate the GRB had to be at high z. 
The $Ly_{\alpha}$ break (12.15 nm) is located between the Y filter (120 nm) and the J filter (213 nm); $z > 7.5$. Using the 
4 filter it is possible, however, to estimate a reasonably good redshift. (Courtesy of N. Tanvir)}\label{fig:grb090423}
\end{center}
\end{figure}

It is only matter to look at the GCN sequence to partially reconstruct the way the two teams arrived at the spectroscopic detection 
of the redshift. While most of the data were kept confidential it was clear that the lack of detection at certain wavelengths and 
the identification of the object at higher wavelengths, we reproduce in Figure\,\ref{fig:grb090423} the best photometric sequence obtained 
with the UKIRT telescope in Hawaii, indicated that we were certainly dealing with a very high z object. In Italy we were able to 
trigger the TNG after we had long discussions about the Nature of the event and after getting the first spectrum (14 hours after 
the burst) we recall that the next day during the Swift teleconference we stressed we were checking the data and preparing a second 
GCN. The two groups had the first evidence that objects, stars producing GRBs, indeed exist at those redshifts and that the search 
for such objects near the dawn of the epoch of the formation of the first stars was fully justified.\\

The host galaxy of GRB\,050904\cite{Berger07} indicate a mass smaller than a few $10^{9}$ solar masses while the metal 
lines\cite{Kawai06} call for a rather low metallicity $Z \sim 0.05 Z_{\odot}$.\\

\begin{figure}
\begin{center}
\includegraphics[width=\hsize]{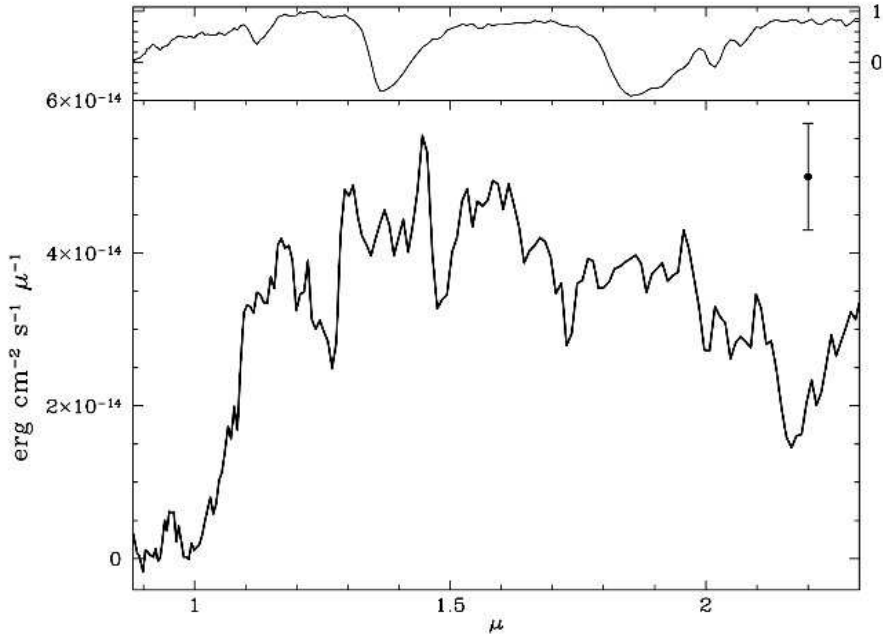}
\caption{Spectrum obtained at the Italian 3.6 m aperture TNG telescope using a de amici prism. On top the standard star that 
shows we are quite sensitive below 100 nm as well. Note that the de mici prism has been designed in order to have a good coverage 
over a large band of NIR wavelength and this is extremely useful in these cases.}\label{fig:grb090423spe}
\end{center}
\end{figure}

Unfortunately the spectrum of GRB\,090423 does not show any detectable emission or absorption line due to the very small signal to 
noise ratio. Our spectrum, obtained about 14 hours since trigger, was obtained with a 3.6 meter telescope while the spectrum obtained 
by the Tanvir group using the VLT (8.2 meter aperture) was gathered at later times in the bands $0.98 - 1.1\,\mu$m and $1.1 - 1.4\,\mu$m respectively. We also attempted to use the X ray spectrum to estimate the metal content of the host galaxy using the absorption edge 
of the X ray spectrum. However these measures, even if indicatives, are rather uncertain both because of the need to disentangle 
various effects (response matrix, variability, power law spectrum etc.) and because the interpretation is not unambiguous 
(absorption systems along the line of sight). The somewhat astonishing result we derive especially from the X ray light curve of 
the afterglow and from the radio observations is that the high z GRB afterglow is completely similar to the long GRBs we observed at 
low redshift.\\

The Radio afterglow\cite{Chandra10} furthermore enable a better estimate of the isotropic (jet) gamma-ray energy 
$E_{\gamma} \sim 10^{53}$ erg ($>2.2\,10^{51}$ erg) and a blast wave kinetic energy $E_{K} \sim 3.8\,10^{53}$ erg 
($> 8.4\,10^{51}$ erg). At high z we seem to detect the very bright end of the GRB luminosity distribution function and it seems 
that the progenitors are the same.

\section{Flares}
The discovery of flares, given the previous observations with Beppo-SAX and the light curves of the first GRBs detected by Swift, 
came to the Swift team as one of the biggest surprises to the extent that we also discussed the possibility they could be an artifact 
of the instrument.\\

\begin{figure}
\begin{center}
\includegraphics[width=\hsize]{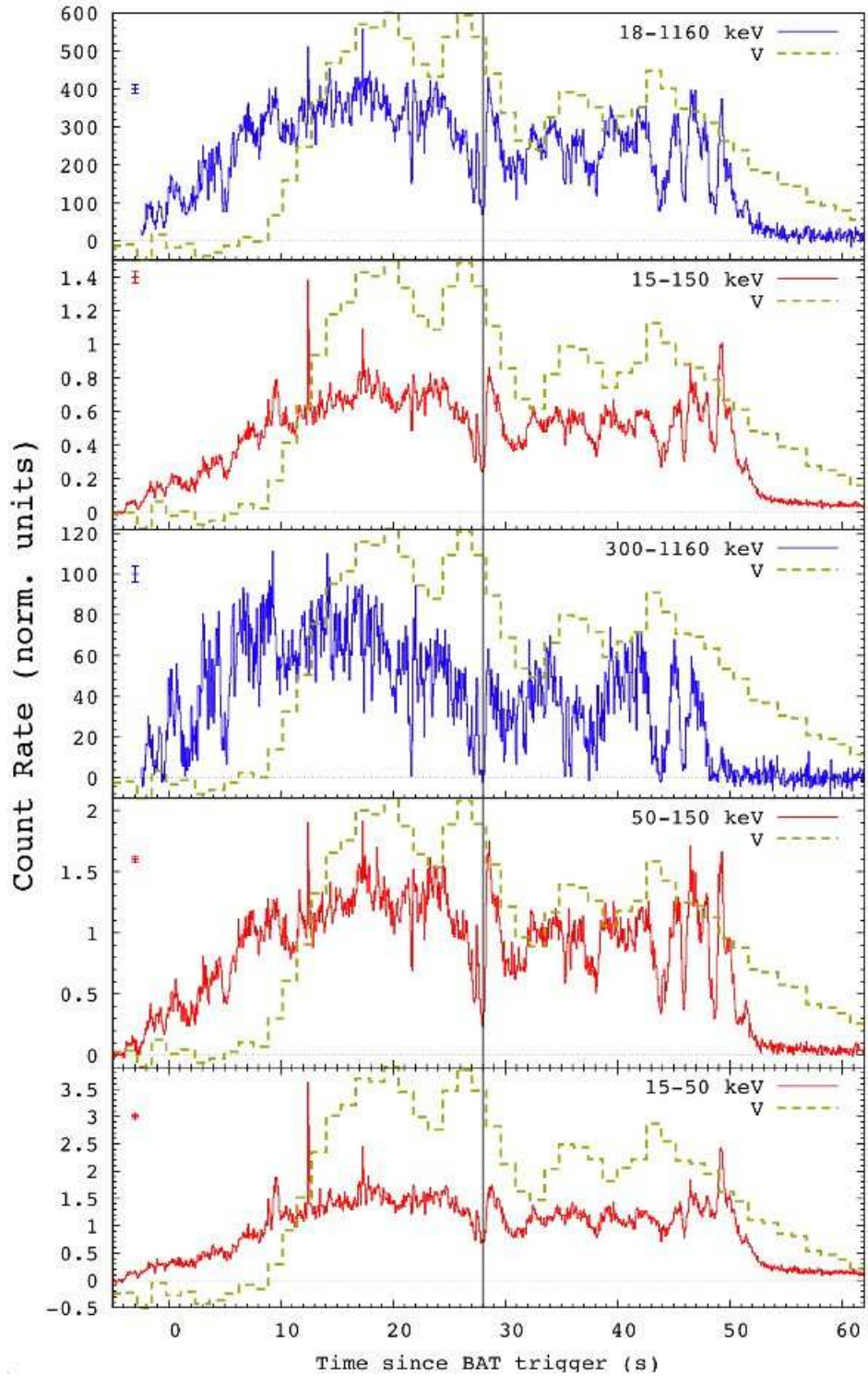}
\caption{The continuous lines in blue and red are the observations obtained by the Konus - Wind satellite and Swift BAT. The yellow 
in the optical by the TORTORA camera mounted on the REM Telescope. Courtesy of C. Guidorzi.}\label{fig:grb080319B}
\end{center}
\end{figure}

Their understanding are of paramount importance because a) they reflect brief and powerful injection of energy that must likely 
be related to the central engine, b) they have been detected both in long and short GRBs indicating a mechanism that must be the 
same in both types of events.  The open questions are many and it is not yet clear whether we have all the observations we need to 
answer them. The observed X-ray light curve shows a number of features that are not fully accounted by the standard 
(internal - external shock) model. The spectrum of the prompt emission may include a thermal component and in any case being rather 
flat below the peak does not fit the synchrotron emission, the afterglow light curve presents an yet unexplained plateau that 
either requires a large injection of energy or may be related to an external shock with very high energy budget however. 
Often (see for instance GRB\,070110\cite{Troja07}) a very steep drop follows that may be eventually related to combined flare 
activity.\\

In GRB\,080319B (the naked eye GRB\cite{Racusin08}) we observed rapid variability both at high energy and in the optical, 
Figure\,\ref{fig:grb080319B}. The optical light curve variability on a scale of 5 to 10 s follows with a delay of about 6 seconds relative 
to the hard X ray variability while differences exist (here however it is essential to plan for higher time resolution at 
optical wavelengths) at higher frequency variability. Even invoking an inverse Compton for the high-energy emission the differences 
in variability and the delay call for a revised model.\\

The short GRBs present in some cases light curves of the afterglow that are similar to those of the long GRBs; in some cases the 
prompt emission present an extended tail that may eventually call for a new classification. The morphology long short is related to 
the general consensus that short GRBs are due to the merging NS+NS and long to the collapse of a massive star. A long lasting 
prompt emission for a short may be at variance with the timing of the phenomenon or with the source of the accreted material. If 
the additional emission during the early phase is due to a left over of the merging\cite{Troja09} then we may have a late accretion 
that should be however present in most of the observed shorts. The afterglow light curve, see for instance GRB\,050724, is often 
similar and presents a steep decay, plateau and further decay. In these cases is the phenomenology exactly the same, environment 
included, independently of the progenitor?\\

\begin{figure}
\begin{center}
\includegraphics[width=\hsize]{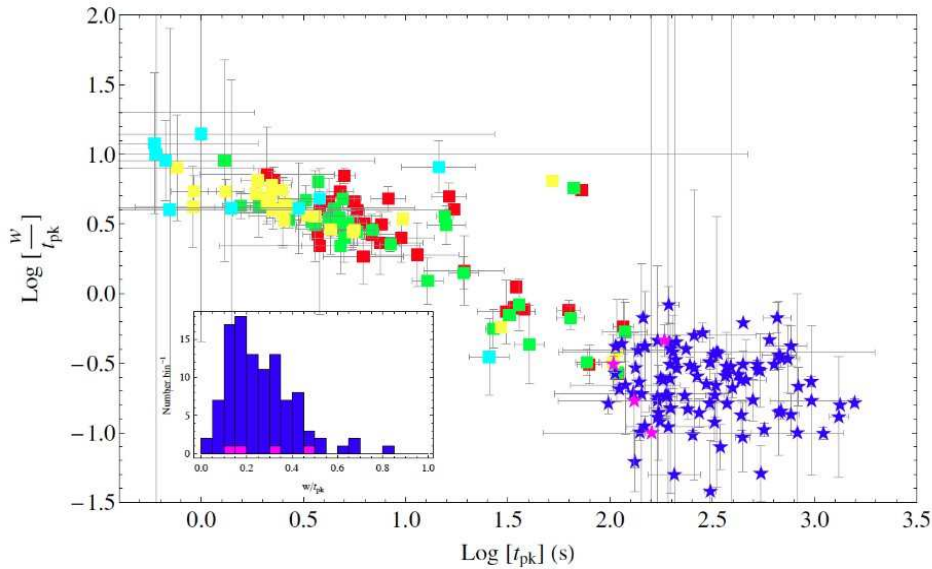}
\caption{Width to $t_{peak}$ ratio as a function of $t_{peak}$, (blue stars). The squares refer to the BATSE pulses measured by 
Ref.\,\protect\refcite{Norris05} and the different colors to different energy channels. In the inset we have the distribution of the ratio with the red 
color referring to the flares observed in shirt GRBs.}\label{fig:tpeak}
\end{center}
\end{figure}

Long and short GRBs present flares in their light curve. This sudden and brief injection of energy that can occur at any time during 
the afterglow may help in determining the mechanism of emission and the characteristics of the central engine. In a simple minded 
model the observed with to peak ratio ($\Delta$ t/t $\sim 0.2$) call for a radius of emission of about $10^{17}$ cm that is too 
large to be produced by internal shock when compared to the deceleration radius, 
$R \sim \left(\frac{E_{K}}{\gamma_{0}\,n\,m_{p}\,c^{2}}\right)^{1/3} \sim 10^{17}$ cm.
We almost completed the analysis of a new sample of 113 X-ray flares, 43 of which have a measured redshift. The sample was analyzed 
in 5 XRT bands (0.3 keV - 10 keV, 0.3 keV - 1 keV, 1 keV - 2 keV, 2 keV - 3 keV and 3 keV - 10 keV) in order to measure some parameters 
as a function of energy. The net result of this analysis\cite{Chincarini10,Margutti10,Bernardini10} is that we confirm 
previous results\cite{Chincarini07,Falcone07} and give more evidence that the flares observed during the afterglow have the 
same characteristics of the pulses observed in the low luminosity (see the distribution function in Ref.\,\refcite{Quilligan02}) GRBs. 
In Figure\,\ref{fig:tpeak} we plot in conjunction of the XRT flares the pulses observed during the prompt emission phase\cite{Norris05}. 
There is continuity with a smooth transition from the prompt emission to the afterglow flares with the difference however that 
while the observed prompt emission pulses maintain a constant width, the width of the XRT flares increase with time.\\
Of outmost importance is the energy budget. The sample is not complete and we did not push the identification for very faint flares 
since one of the constraints we had in the sample was to be able to measure the profile.  Flares are very energetic and represent 
bursts of energy of about $10\%$ the energy the energy emitted during the afterglow. In a few cases, GRB\,050502B is the prototype, 
the energy emitted during the flare is about the whole energy emitted during the afterglow. Late flares show a smaller amplitude 
and a much larger width, the energy however does not change that much even if it is somewhat smaller that the energy emitted 
by early flares\cite{Bernardini10}.\\

\begin{figure}
\begin{center}
\includegraphics[width=\hsize]{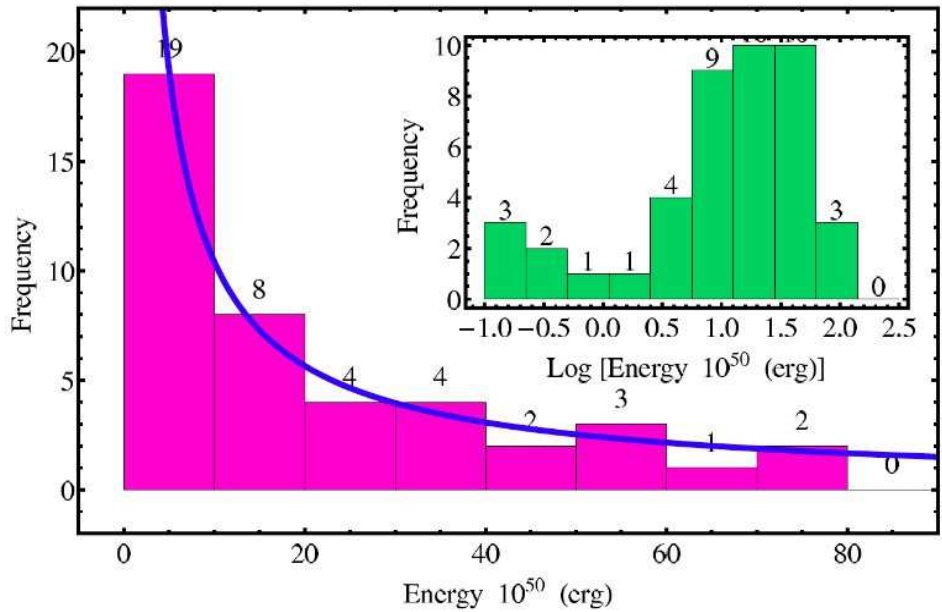}
\caption{Distribution of energy of flares having spectroscopic redhift. In the histogram of the inset the energy has been 
expressed in log units.}\label{fig:energy}
\end{center}
\end{figure}

As for the pulses observed during the prompt emission we measure a variation of the width with energy. In this case we define an 
efficient energy $E_{eff}$ as the energy weighted over the pass band by the following relation:

\begin{equation}
E_{eff} = (1 + z) \frac{\int E\,f(E)\,RM\,dE}{\int f(E)\,RM\,dE}
\end{equation}

In this way and after correcting and using the rest frame width we find a net correlation between the width and the energy of 
the flare: at higher energies the flares are narrower.\\

\begin{figure}
\begin{center}
\includegraphics[width=\hsize]{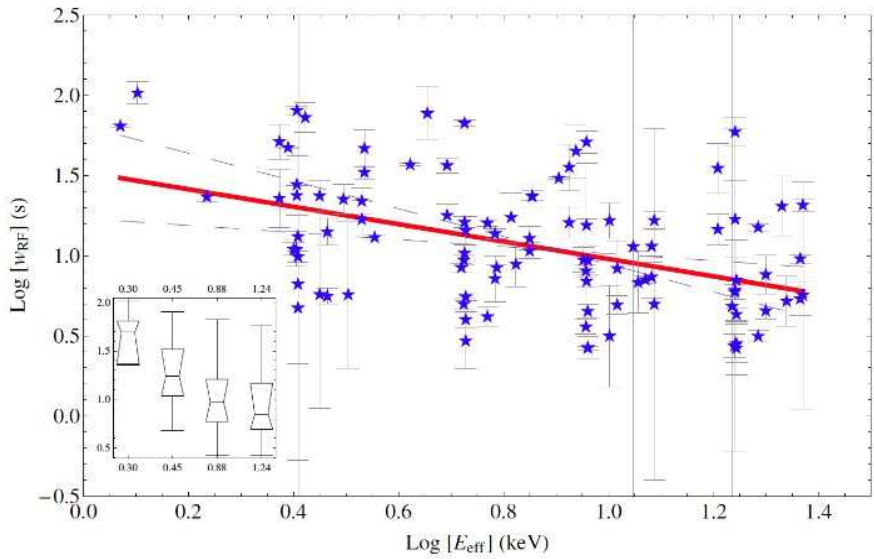}
\caption{Plot of the rest frame width as a function of the efficient energy. In the inset the related quantile plot.}\label{fig:energy1}
\end{center}
\end{figure}

Indeed the correlation we find: $w_{RF} = 10^{1.5} E^{-0.5}_{eff}$ is very similar to that estimated by 
Ref.\,\refcite{Fenimore95} for pulses observed by BATSE: $w \propto E^{-0.4}$.\\

\textbf{It is remarkable that in all the correlations the flares reflect the same characteristics of the prompt emission pulses indicating a similar mechanism at work and likely the same origin so that this activity lasts from the prompt emission to about a million second after the trigger}.

\section{Discussion and conclusions}
Recent observations challenge the fireball internal shock model. Optical observations do not show strong evidence of the reverse 
shock\cite{Kumar07}, and in any case even if present does not seem to be one of the main component of the early optical 
emission\cite{Oates09}. The reverse shock is weak or suppressed in the magneto hydro dynamical models of 
GRBs\cite{Thompson94,Spruit01,Lyutikov03} so that the goal is to find way to estimate whether
$\sigma_{0}=\frac{F_{p}}{F_{b}}=\frac{B_{0}^{2}}{(4\,\pi\,\gamma_{0}\,\rho\,c^{2})}$ is $\gg 1$ (the jet is dominated by 
a Pointing flux) or $\ll 1$ (the kinetic energy of the baryonic jet dominates).\\

\begin{figure}
\begin{center}
\includegraphics[width=\hsize]{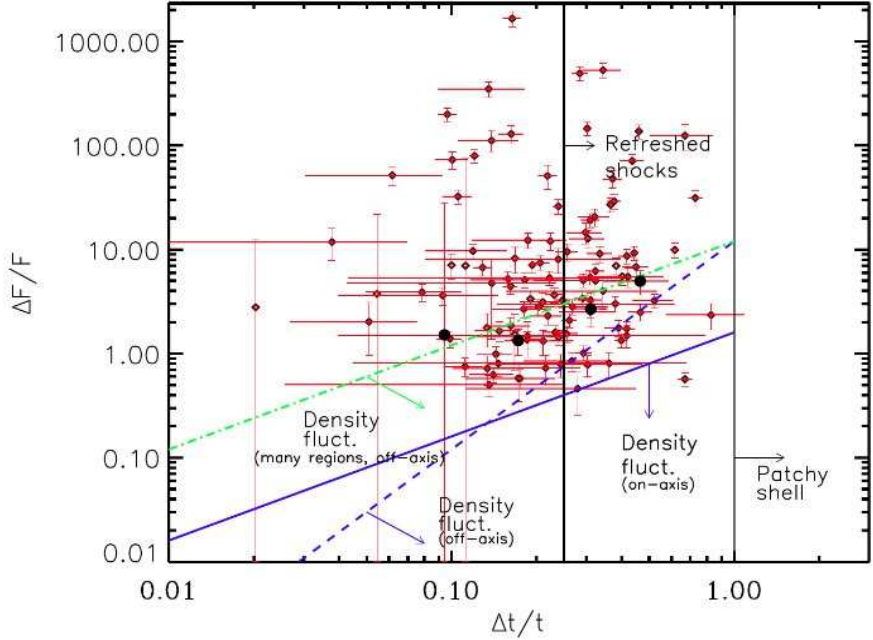}
\caption{The relative variability of the flares as a function of the width to time ratio has been plotted. The limits shown by the 
various lines have been calculated according to the equations given in Ref.\,\protect\refcite{Ioka05}. The 4 black dots refer to short GRBs.}\label{fig:variability}
\end{center}
\end{figure}

We should add incidentally, that the mechanism by which strong magnetic fields and the acceleration of particles to very high energy 
is achieved is not yet fully known. It needs to be demonstrated that the linear Fermi mechanism is capable of doing the job following 
the back and for transit of the particles in the shock region.\\

To summarize part of the results in a diagram we refer to plot of the ratio $\Delta F/F$ versus $\Delta t/t$, Figure\,\ref{fig:variability}.
Here $\Delta F$ is the variation of the flux above the underlying continuum over the flux observed in the underlying continuum and
$\Delta t/t$ the flare width over the time of the peak of the flare.\\

The complete absence of flares with $\Delta t/t > 1$ confirms that flares cannot be due to patchy shells. Flares furthermore seem 
to have an high probability, as expected, to be observed off axis and a large number of flares would, according to this plot, 
agree with the refreshed shock model.\\

Flares are not necessarily the result of late central engine activity, but may be produced in the decelerating phase of the flow. 
High $\sigma_{0}$ values may lead to MHD instabilities during the interaction with the interstellar medium. In this case the 
isotropic equivalent energy emitted in a single flare and produced by a single reconnection event, is related to the ratio 
width/$t_{peak}$ by the relation:

\begin{equation}
E_{flare} \leq 5\,\epsilon \left(\frac{width}{t_{peak}}\right)^{3} \frac{E_{forward\,shock}}{\alpha^{2}}
\end{equation}

where $\epsilon \sim 0.1$ and $\alpha$ in the range 2 - 4 depending on the density of the ISM. The flares we observe 
are not in contradiction within errors with this relation but, at the same time, they do not prove it.\\
In conclusion we find that flares seem to retain memory of the previous event so that, as time progresses, each 
flare is weaker and softer of the preceding one. And finally while we are making significant progress in 
characterizing these events observationally, at the moment there is no satisfactory model explaining their origin, 
evolution and energetics.

\section*{Acknowledgments}
This work is supported by ASI grant SWIFT I/011/07/0, by the Ministry of University and Research of 
Italy (PRIN, MIUR, 2007TNYZXL), by MAE and by the University of Milano Bicocca (Italy).

\end{document}